% hep-th/0402135, v1
% last edited by AS, 2/17/04

\input harvmac

\noblackbox

\def\IZ{\relax\ifmmode\mathchoice
{\hbox{\cmss Z\kern-.4em Z}}{\hbox{\cmss Z\kern-.4em Z}}
{\lower.9pt\hbox{\cmsss Z\kern-.4em Z}}
{\lower1.2pt\hbox{\cmsss Z\kern-.4em Z}}\else{\cmss Z\kern-.4em Z}\fi}
\def\IB{\relax{\rm I\kern-.18em B}}
\def\IC{{\relax\hbox{\kern.3em{\cmss I}$\kern-.4em{\rm C}$}}}
\def\ID{\relax{\rm I\kern-.18em D}}
\def\IE{\relax{\rm I\kern-.18em E}}
\def\IF{\relax{\rm I\kern-.18em F}}
\def\IG{\relax\hbox{$\inbar\kern-.3em{\rm G}$}}
\def\IGa{\relax\hbox{${\rm I}\kern-.18em\Gamma$}}
\def\IH{\relax{\rm I\kern-.18em H}}
\def\II{\relax{\rm I\kern-.18em I}}
\def\IK{\relax{\rm I\kern-.18em K}}
\def\IP{\relax{\rm I\kern-.18em P}}
%\def\IX{\relax{\rm X\kern-.01em X}}
%this doesn't work

\font\cmss=cmss10 \font\cmsss=cmss10 at 7pt
\def\IR{\relax{\rm I\kern-.18em R}}

\def\frac#1#2{{#1 \over #2}}

\def\OL#1{ \kern1pt\overline{\kern-1pt#1
     \kern-1pt}\kern1pt }

%\AndrianopoliSA
\lref\ferrara{ L.~Andrianopoli, S.~Ferrara and M.~Trigiante, ``Fluxes,
supersymmetry breaking and gauged
supergravity,'' arXiv:hep-th/0307139.
%%CITATION = HEP-TH 0307139;%%
}

%\MayrHH
\lref\MayrHH{ P.~Mayr,
  ``On supersymmetry breaking in string theory and its realization in brane
%worlds,''
Nucl.\ Phys.\ B {\bf 593}, 99 (2001) [arXiv:hep-th/0003198].
%%CITATION = HEP-TH 0003198;%%
}

%\TaylorII
\lref\TaylorII{ T.~R.~Taylor and C.~Vafa, ``RR flux on Calabi-Yau and
partial supersymmetry breaking,'' Phys.\
Lett.\ B {\bf 474}, 130 (2000) [arXiv:hep-th/9912152].
%%CITATION = HEP-TH 9912152;%%
}

%\GiryavetsVD
\lref\GiryavetsVD{ A.~Giryavets, S.~Kachru, P.~K.~Tripathy and
S.~P.~Trivedi, ``Flux compactifications on
Calabi-Yau threefolds,'' arXiv:hep-th/0312104.
%%CITATION = HEP-TH 0312104;%%
}

%\TripathyQW
\lref\TripathyQW{
P.~K.~Tripathy and S.~P.~Trivedi,
``Compactification with flux on K3 and tori,'' JHEP {\bf 0303}, 028 (2003)
[arXiv:hep-th/0301139].
%%CITATION = HEP-TH 0301139;%%
}

\lref\GVW{S. Gukov, C. Vafa and E. Witten, ``CFTs from Calabi-Yau
Fourfolds,'' Nucl. Phys. {\bf B584}, 69
(2000)[arXiv:hep-th/9906070]}

\lref\GKP{ S.~B.~Giddings, S.~Kachru and J.~Polchinski,` `Hierarchies from
fluxes in string compactifications,''
Phys.\ Rev.\ D {\bf 66}, 106006 (2002) [arXiv:hep-th/0105097].
%%CITATION = HEP-TH 0105097;%%
}

\lref\KKLT{ S.~Kachru, R.~Kallosh, A.~Linde and S.~P.~Trivedi,``De Sitter
vacua in string theory,'' Phys.\ Rev.\
D {\bf 68},046005 (2003) [arXiv:hep-th/0301240].
%%CITATION = HEP-TH 0301240;%%
}

\lref\KST{ S.~Kachru, M.~B.~Schulz and S.~Trivedi, ``Moduli stabilization
from fluxes in a simple IIB
orientifold,'' JHEP {\bf0310}, 007 (2003) [arXiv:hep-th/0201028].
%%CITATION = HEP-TH 0201028;%%
}

%\KachruNS
\lref\KachruNS{ S.~Kachru, X.~Liu, M.~B.~Schulz and S.~P.~Trivedi,
``Supersymmetry changing bubbles in string
theory,'' JHEP {\bf 0305}, 014 (2003) [arXiv:hep-th/0205108].
%%CITATION = HEP-TH 0205108;%%
}

%\FreyHF
\lref\FreyHF{ A.~R.~Frey and J.~Polchinski, ``N = 3 warped
compactifications,'' Phys.\ Rev.\ D {\bf 65}, 126009
(2002) [arXiv:hep-th/0201029].
%%CITATION = HEP-TH 0201029;%%
}

\lref\WSuper{E.~Witten,``Non-Perturbative Superpotentials In String
Theory,'' Nucl.\ Phys.\ B {\bf 474}, 343
(1996) [arXiv:hep-th/9604030].
%%CITATION = HEP-TH 9604030;%%
}

\lref\Candelas{ P.~Candelas and X.~de la Ossa,``Moduli Space Of Calabi-Yau
Manifolds,'' Nucl.\ Phys.\ B {\bf
355}, 455 (1991).
%%CITATION = NUPHA,B355,455;%%
}

\lref\mathematica{ S. Wolfram, ``The Mathematic Book,'' {\it Wolfram
Media/Cambridge Univ. Pr., 1999} }

\lref\JoeTwo{ J.~Polchinski, ``String Theory. Vol. 2: Superstring Theory And
Beyond,'' {\it  Cambridge, UK:
Univ. Pr. (1998) 531 p}.}

\font\fivebf  =cmbx10  scaled 500 % five point bold
     \font\sevenbf =cmbx10  scaled 700 % seven point bold
     \font\tenbf   =cmbx10             % ten point bold
     \font\fivemb  =cmmib10 scaled 500 % five point math bold
     \font\sevenmb =cmmib10 scaled 700 % seven point math bold
     \font\tenmb   =cmmib10            % ten point math bold

\def\boldmath{\textfont0=\tenbf           \scriptfont0=\sevenbf
                \scriptscriptfont0=\fivebf  \textfont1=\tenmb
                \scriptfont1=\sevenmb       \scriptscriptfont1=\fivemb}

\lref\MSS{
%\MaloneyRR
A.~Maloney, E.~Silverstein and A.~Strominger, ``De Sitter space in
noncritical string theory,''
arXiv:hep-th/0205316;
%%CITATION = HEP-TH 0205316;%%
%\SilversteinXN
E.~Silverstein, ``(A)dS backgrounds from asymmetric orientifolds,''
arXiv:hep-th/0106209.
%%CITATION = HEP-TH 0106209;%%
}

\lref\trap{L. Kofman, A. Linde, X. Liu, A. Maloney, L. McAllister, and E.
Silverstein, ``Moduli Trapping from
Particle Production", to appear.}

\lref\dine{M. Dine, ``Towards a Solution of the Moduli Problems of String
Cosmology,'' Phys.Lett. {\bf{B}}482
(2000) 213, hep-th/0002047 \semi M. Dine, Y. Nir, and Y. Shadmi, ``Enhanced
Symmetries and the Ground State of
String Theory,'' Phys.Lett. {\bf{B}}438 (1998) 61, hep-th/9806124.}

%\BurgessIC
\lref\renata{ C.~P.~Burgess, R.~Kallosh and F.~Quevedo, ``de Sitter string
vacua from supersymmetric D-terms,''
JHEP {\bf 0310}, 056 (2003) [arXiv:hep-th/0309187].
%%CITATION = HEP-TH 0309187;%%
}

%\SilversteinJP
\lref\SilversteinJP{ E.~Silverstein, ``AdS and dS entropy from string
junctions,'' arXiv:hep-th/0308175.
%%CITATION = HEP-TH 0308175;%%
}

%\FabingerGP
\lref\FabingerGP{ M.~Fabinger and E.~Silverstein, ``D-Sitter space: Causal
structure, thermodynamics, and
entropy,'' arXiv:hep-th/0304220.
%%CITATION = HEP-TH 0304220;%%
}

%\SusskindKW
\lref\SusskindKW{ L.~Susskind, ``The anthropic landscape of string theory,''
arXiv:hep-th/0302219.
%%CITATION = HEP-TH 0302219;%%
}

%\AshokGK
\lref\AshokGK{ S.~Ashok and M.~R.~Douglas, ``Counting flux vacua,''
arXiv:hep-th/0307049.
%%CITATION = HEP-TH 0307049;%%
}

\lref\achar{
%\AcharyaKV
B.~S.~Acharya, ``A moduli fixing mechanism in M theory,''
arXiv:hep-th/0212294.
%%CITATION = HEP-TH 0212294;%%
}

\lref\W{
%\WittenQJ
E.~Witten, ``Anti-de Sitter space and holography,'' Adv.\ Theor.\ Math.\
Phys.\  {\bf 2}, 253 (1998)
[arXiv:hep-th/9802150].
%%CITATION = HEP-TH 9802150;%%
}

%\StromingerSH
\lref\StromingerSH{ A.~Strominger and C.~Vafa, ``Microscopic Origin of the
Bekenstein-Hawking Entropy,'' Phys.\
Lett.\ B {\bf 379}, 99 (1996) [arXiv:hep-th/9601029].
%%CITATION = HEP-TH 9601029;%%
}

\lref\GKPads{
%\GubserBC
S.~S.~Gubser, I.~R.~Klebanov and A.~M.~Polyakov, ``Gauge theory correlators
from non-critical string theory,''
Phys.\ Lett.\ B {\bf 428}, 105 (1998) [arXiv:hep-th/9802109].
%%CITATION = HEP-TH 9802109;%%
}

%\KachruNS
\lref\KachruNS{ S.~Kachru, X.~Liu, M.~B.~Schulz and S.~P.~Trivedi,
``Supersymmetry changing bubbles in string
theory,'' JHEP {\bf 0305}, 014 (2003) [arXiv:hep-th/0205108].
%%CITATION = HEP-TH 0205108;%%
}

\lref\GH{
%\GibbonsMU
G.~W.~Gibbons and S.~W.~Hawking, ``Cosmological Event Horizons,
Thermodynamics, And Particle Creation,'' Phys.\
Rev.\ D {\bf 15}, 2738 (1977).
%%CITATION = PHRVA,D15,2738;%%
}

\lref\DS{
%\FabingerGP
M.~Fabinger and E.~Silverstein, ``D-Sitter space: Causal structure,
thermodynamics, and entropy,''
arXiv:hep-th/0304220.
%%CITATION = HEP-TH 0304220;%%
}

\lref\juan{
%\MaldacenaRE
J.~M.~Maldacena, ``The large N limit of superconformal field theories and
supergravity,'' Adv.\ Theor.\ Math.\
Phys.\ {\bf 2}, 231 (1998) [Int.\ J.\ Theor.\ Phys.\  {\bf 38}, 1113 (1999)]
[arXiv:hep-th/9711200].
%%CITATION = HEP-TH 9711200;%%
}

\lref\dSCFT{
%\StromingerPN
A.~Strominger, ``The dS/CFT correspondence,'' JHEP {\bf 0110}, 034 (2001)
[arXiv:hep-th/0106113]
%%CITATION = HEP-TH 0106113;%%
}

%\AcharyaKV
\lref\AcharyaKV{ B.~S.~Acharya, ``A moduli fixing mechanism in M theory,''
arXiv:hep-th/0212294.
%%CITATION = HEP-TH 0212294;%%
}

\lref\dSObj{
%\WittenKN
E.~Witten, ``Quantum gravity in de Sitter space,'' arXiv:hep-th/0106109;
%%CITATION = HEP-TH 0106109;%%
%\FischlerYJ
W.~Fischler, A.~Kashani-Poor, R.~McNees and S.~Paban, ``The acceleration of
the universe, a challenge for string
theory,'' JHEP {\bf 0107}, 003 (2001) [arXiv:hep-th/0104181];
%%CITATION = HEP-TH 0104181;%%
%\HellermanYI
S.~Hellerman, N.~Kaloper and L.~Susskind, ``String theory and
quintessence,'' JHEP {\bf 0106}, 003 (2001)
[arXiv:hep-th/0104180];
%%CITATION = HEP-TH 0104180;%%
%\GoheerVF
N.~Goheer, M.~Kleban and L.~Susskind, ``The trouble with de Sitter space,''
JHEP {\bf 0307}, 056 (2003)
[arXiv:hep-th/0212209].
%%CITATION = HEP-TH 0212209;%%
%\DysonNT
L.~Dyson, J.~Lindesay and L.~Susskind, ``Is there really a de Sitter/CFT
duality,'' JHEP {\bf 0208}, 045 (2002)
[arXiv:hep-th/0202163];
%%CITATION = HEP-TH 0202163;%%
%\BanksWR
T.~Banks, W.~Fischler and S.~Paban, ``Recurrent nightmares?: Measurement
theory in de Sitter space,'' JHEP {\bf
0212}, 062 (2002) [arXiv:hep-th/0210160];
%%CITATION = HEP-TH 0210160;%%
%\BanksYP
T.~Banks and W.~Fischler, ``M-theory observables for cosmological
space-times,'' arXiv:hep-th/0102077.
%%CITATION = HEP-TH 0102077;%%
}

\lref\stringstalk{E. Silverstein, talk at Strings 2003}

\lref\us{M. Fabinger, S. Hellerman, E. Silverstein, and others, in
progress.}

\lref\KLT{
%\KrausHV
P.~Kraus, F.~Larsen and S.~P.~Trivedi, ``The Coulomb branch of gauge theory
from rotating branes,'' JHEP {\bf
9903}, 003 (1999) [arXiv:hep-th/9811120].
%%CITATION = HEP-TH 9811120;%%
}

%\AbbottQF
\lref\AbbottQF{ L.~F.~Abbott, ``A Mechanism For Reducing The Value Of The
Cosmological Constant,'' Phys.\ Lett.\
B {\bf 150}, 427 (1985).
%%CITATION = PHLTA,B150,427;%%
}

%\BanksMB
\lref\BanksMB{ T.~Banks, M.~Dine and N.~Seiberg, ``Irrational axions as a
solution of the strong CP problem in
an eternal universe,'' Phys.\ Lett.\ B {\bf 273}, 105 (1991)
[arXiv:hep-th/9109040].
%%CITATION = HEP-TH 9109040;%%
}

%\DasguptaSS
\lref\DasguptaSS{ K.~Dasgupta, G.~Rajesh and S.~Sethi, ``M theory,
orientifolds and G-flux,'' JHEP {\bf 9908},
023 (1999) [arXiv:hep-th/9908088].
%%CITATION = HEP-TH 9908088;%%
}

%\BrownKG
\lref\BrownKG{ J.~D.~Brown and C.~Teitelboim, ``Neutralization Of The
Cosmological Constant By Membrane
Creation,'' Nucl.\ Phys.\ B {\bf 297}, 787 (1988).
%%CITATION = NUPHA,B297,787;%%
}

%\BoussoXA
\lref\BoussoXA{ R.~Bousso and J.~Polchinski,
   %``Quantization of four-form fluxes and dynamical neutralization of the
%cosmological constant,''
JHEP {\bf 0006}, 006 (2000) [arXiv:hep-th/0004134].
%%CITATION = HEP-TH 0004134;%%
}

%\FengIF
\lref\FengIF{ J.~L.~Feng, J.~March-Russell, S.~Sethi and F.~Wilczek,
``Saltatory relaxation of the cosmological
constant,'' Nucl.\ Phys.\ B {\bf 602}, 307 (2001) [arXiv:hep-th/0005276].
%%CITATION = HEP-TH 0005276;%%
}

%\KachruGS
\lref\KachruGS{ S.~Kachru, J.~Pearson and H.~Verlinde,
   ``Brane/flux annihilation and the string dual of a non-supersymmetric
field
%theory,''
JHEP {\bf 0206}, 021 (2002) [arXiv:hep-th/0112197].
%%CITATION = HEP-TH 0112197;%%
}

\lref\kls{kls refs}

\lref\gs{G. Dvali and S. Kachru, ``New Old Inflation,''
hep-th/0309095.}

\lref\SilversteinHF{E. Silverstein and D. Tong, ``Scalar Speed
Limits and Cosmology: Acceleration from D-cceleration,''
hep-th/0310221.}

\lref\jarv{L. Jarv, T. Mohaupt, and F. Saueressig, ``M-theory
cosmologies from singular Calabi-Yau compactifications,''
hep-th/0310174.}

\lref\bd{N. D. Birrell and P.C.W. Davies, {\it Quantum Fields in
Curved Space}, Cambridge University Press, Cambridge, England
(1982).}

\lref\horne{J. Horne and G. Moore, ``Chaotic Coupling Constants,''
   Nucl.Phys. {\bf{B}}432 (1994) 109, hep-th/9403058.}

\lref\hyb{A. Linde, ``Hybrid Inflation,'' Phys.Rev. {\bf{D}}49
(1994) 748, astro-ph/9307002.}

\lref\sw{N. Seiberg and E. Witten, ``Electric-Magnetic Duality,
Monopole Condensation, and Confinement in N=2 Supersymmetric
Yang-Mills Theory,'' Nucl.Phys. {\bf{B}}426 (1994) 19,
hep-th/9407087 \semi K. Intriligator and N. Seiberg, ``Lectures on
Supersymmetric Gauge Theories and Electric-Magnetic Duality,''
Nucl.Phys.Proc.Suppl. 45BC (1996) 1, hep-th/9509066.}

\lref\flux{some papers on moduli stabilization from fluxes}

%\draft

\Title{\vbox{\baselineskip12pt\hbox{hep-th/0402135}
\hbox{SLAC-PUB-10347}\hbox{SU-ITP-04/06}}} {\vbox{
\centerline{The Scaling of the No-Scale Potential}
\bigskip
\centerline{and de Sitter Model Building} }}

\centerline{Alex Saltman and Eva Silverstein \footnote{$^*$} {SLAC and
Department of Physics, Stanford
University, Stanford, CA 94309}}

\vskip .3in We propose a variant of the KKLT (A)dS flux vacuum construction which does not require an antibrane
to source the volume modulus.  The strategy is to find nonzero local minima of the no-scale potential in the
complex structure and dilaton directions in moduli space.  The corresponding no-scale potential expanded about
this point sources the volume modulus in the same way as does the antibrane of the KKLT construction.  We
exhibit explicit examples of such nonzero local minima of the no-scale potential in a simple toroidal
orientifold model.

\smallskip
\Date{}
%\listtoc
%\writetoc

%\vfill \eject

\newsec{Introduction}

De Sitter and anti-de Sitter vacua of string theory that fix all moduli were constructed recently
\refs{\MSS,\KKLT,\AcharyaKV}. The de Sitter models offer the prospect of modelling dark energy and inflation in
string theory, as well as providing a concrete setting in which to analyze basic quantum gravitational issues
(such as the de Sitter entropy \refs{\SilversteinJP,\FabingerGP} and the rich cosmology and particle physics of
multiple vacua \refs{\BoussoXA,\SusskindKW,\AshokGK}).

The KKLT models \KKLT\ may be of particular interest as they arise in a context with low-energy supersymmetric
effective field theory and naturally incorporate other appealing model-building features, such as warping to
obtain small numbers.  In this paper we will provide explicit examples of a simple variant of the KKLT
construction.

One ingredient of the model \KKLT\ is an anti-D3-brane in a warped 3-brane
throat of a Calabi-Yau
compactification of type IIB string theory \KachruGS. This produces a
contribution to the four-dimensional
Einstein-frame potential for the overall volume $V_{CY}$ of the Calabi-Yau
\eqn\volscale{V_{\overline{D3}}\sim {1\over V_{CY}^2}\tilde V}
where $\tilde V$ does not depend on the overall volume, to a first
approximation, and is tunably small if the
${\overline{D3}}$ sits at the bottom of a sufficiently warped throat. The
contribution \volscale\ can be
understood simply as arising from the $\overline{D3}$ tension in
four-dimensional Einstein frame. It is not yet
precisely known how to package the contributions of the
$\overline{D3}$-brane in terms of the superpotential,
K\"ahler potential, and D terms of a low energy supersymmetric effective
field theory.  (Recently, an
interesting variant of KKLT has been proposed in which the antibrane is
embedded in a 7-brane, giving a D-term
realization of it \renata; it will be interesting to see if explicit tadpole
and tachyon free models arise also
in that context.)

The potential energy of a generic model can be written in ${\cal N}=1$
supersymmetric form in terms of a K\"ahler potential $K$ and superpotential
$W$ as
\eqn\Vsup{V = e^{K}(\sum_{a,b} g^{a \bar b} D_a W \overline{D_b W} - 3
|W|^2)+{\rm D ~ terms}}
where $a$ and $b$ run over all moduli and the axion-dilaton and $g_{a \bar
b} = \partial_a \partial_{\bar b} K$ is
the metric on moduli space. In a no-scale model, such as those developed in
\GKP, this reduces to
\eqn\noscale{V_{\rm no-scale} = e^{K}(\sum_{i,j} g^{i \bar j} D_i W
\overline{D_j W})}
where $i$ and $j$ run over only complex structure moduli and the
axion-dilaton (collectively $\Phi_i$), the
K\"ahler and D3-brane position moduli having cancelled out. Thus far,
solutions to this potential have been
found \refs{\KKLT,\GKP,\KST,\FreyHF,\TripathyQW}, by solving $D_i W=0$,
$\forall ~i$, which produces a
zero-energy minimum of \noscale\ with fixed complex structure moduli. The
K\"ahler moduli only appear in
\noscale\ in the prefactor $e^K$, so they are fixed by other ingredients--a
balance of forces from \volscale\
and two terms arising from no-scale breaking effects ($\alpha^\prime$
corrections and nonperturbative effects).
This leaves the complex moduli close to their original positions as
stabilized by the no-scale potential.

However, the functional form of \noscale\ immediately suggests an
interesting alternative to this construction.
The overall volume appears as a K\"ahler modulus in a chiral multiplet
$\rho$ whose imaginary part is
$V_{CY}^{2/3}/g_s$, and $e^K\propto 1/({\rm Im} \rho)^3 \propto 1/V_{CY}^2$.
So if there
is a minimum of the potential in
the complex-structure and axion-dilaton directions at {\it nonzero} $V_{\rm
no-scale}$,
\eqn\conditions{\del_i V |_{\Phi_i^*} = 0 ~~~~~~~~ V(\Phi_i^*)>0, }
with the eigenvalues of $\del_i\del_j V$ positive, it provides a source for
$\rho$ which is identical in form to
that provided by the antibrane \volscale.

In a similar manner to \KKLT, we can play this positive-power-law potential \volscale\ against contributions
coming from no-scale-violating effects.  For example, it is straightforward to expand the potential \Vsup\ by
taking $W\to W_{ns}+\Delta W$ where $\Delta W$ is non-perturbatively small in the volume modulus and $W_{ns}$ is
the no-scale superpotential (independent of the volume). Combining the resulting terms with our positive
potential, we obtain contributions scaling like $\sqrt{\tilde V}\Delta W/({\rm Im} \rho)^3$ and $\Delta W/({\rm
Im} \rho)^4$ along with the ${\tilde V}/({\rm Im} \rho)^3$ from \volscale; with appropriate signs, if $\tilde V$
is small, these terms can play off each other to fix the volume at a large value. In the example of \KKLT, the
small number required is obtained by positioning the antibrane at the bottom of a warped throat.  In our case,
we will have to rely on the richness of the set of possible fluxes to obtain a small number.

This class of models is in some sense a natural completion of that offered by KKLT, since there are domain walls
coming from wrapped fivebranes on which the antibrane (as well as other threebranes) can end.  So the set of
models without an antibrane is physically connected to the original set of KKLT models through bubble nucleation
\refs{\BrownKG,\BoussoXA,\KachruNS}, and via the deformation of the system onto its approximate Coulomb branch
\refs{\FabingerGP,\SilversteinJP}.

Our strategy has the advantage that the supersymmetric effective
field theory description is clear, and shows that
there is no a priori need for warped throats. In
realizing our idea with explicit examples, we will work with toroidal
orientifolds as in
\refs{\DasguptaSS,\KST,\FreyHF}\ rather than using generic warped
Calabi-Yau manifolds. As in the early work on string
compactification phenomenology, there is a practical trade-off
between genericity and calculability.  We will also find it very
useful to work near a locus of enhanced symmetry within the
toroidal orientifold model's moduli space. Again this entails some
loss of generality, but on the other hand there are various
interesting physical mechanisms and scenarios which favor
symmetries (such as \refs{\trap,\dine}\ and avoidance of proton
decay).

\newsec{A Concrete Model:  De-Sitter Vacua of the $\boldmath{T^6 / {\IZ_2}}$
Orientifold}

In the KKLT models, before $\alpha^\prime$ and nonperturbative
correction, the potential takes the form \noscale, with K\"ahler potential
\eqn\kahlerpot{K = -3 \ln[-i(\rho - \bar \rho)] - \ln[-i(\phi - \bar \phi)]
- \ln[-i\int_M \Omega
\wedge {\bar \Omega}]}
and superpotential \refs{\GVW,\TaylorII,\MayrHH}
\eqn\superpot{W = \int_M G_{3} \wedge \Omega.}
Here $\phi$ is the IIB axiodilaton (a departure from the notation of \KKLT) and $\rho$ is the superfield with
imaginary part $({\rm Im}~\phi) e^{4u}$, where $e^u$ scales with the linear size of the Calabi-Yau. Thus the
$\rho$ dependence of the potential is simply a multiplicative factor of $1/ ({\rm Im} ~\rho)^3$ from the K\"ahler
potential as mentioned in \S1.

The potentials \noscale\ that one obtains from generic Calabi-Yau manifolds
have a rich structure, and though
the moduli spaces are of high dimension, we expect them to produce
metastable solutions of \conditions. But, for
specificity, we now specialize to the case of flux compactifications of type
IIB on the $T^6/\IZ_2$ orientifold
\KST\FreyHF.

\subsec{The $T^6/\IZ_2$ geometry}

In fact this naively simpler case has several complications of its own
coming from the fact that there is no
preferred decomposition into K\"ahler and complex structure moduli.  (This
can be thought of in spacetime
language as coming from the fact that the UV theory has ${\cal N}=4$
supersymmetry in four dimensions.) In
particular, the derivation of the no-scale potential and flux superpotential
\noscale\superpot\ breaks down in
this case. Ultimately we will simply calculate the potential energy in
components and minimize it, exhibiting
explicit tachyon free models in a way which bypasses these subtleties.

In order to calculate potentials for moduli in various flux configurations,
we introduce a convenient set
of real global coordinates on $T^6$: $x^i$, $y^i$, $i=1,...,3$ with the
identifications $x^i \equiv x^i+ 1$,
$y^i \equiv y^i + 1$. Choices of complex structure can be parameterized by
complex numbers $\tau^{ij}$,
$i,j=1,...,3$ such that
$$z^i = x^i + \tau^{ij} y^j$$
are global holomorphic coordinates. In these coordinates, the explicit
orientifold is $T^6 / (\Omega R
(-1)^{F_L})$ where $R$: $(x^i,y^i) \rightarrow -(x^i,y^i)$. The holomorphic
three-form can be taken to be
\eqn\holo{\Omega = dz^1 \wedge dz^2 \wedge dz^3}
and the metric can be chosen to be
\eqn\Tmetric{ds^2= dz^id\bar z^{\bar i}}

In this parameterization, deformations of $\tau^{ij}$ induce deformations of
the metric that are a combination
of K\"ahler ($\delta g_{i\bar j}\ne 0$) and complex structure ($\delta
g_{ij}\ne 0$) perturbations.\foot{It
should be possible to correct the metric with some dependence on $\tau^{ij}$
that removes the K\"ahler
perturbations, as is commonly done on $T^2$ by defining the K\"ahler form as
$1/(\bar \tau - \tau) dz  d \bar
z$. However, for simplicity, we will not make such a choice.} In addition
(independent of the parameterization),
some deformations of $\tau^{ij}$ induce no complex-structure metric
perturbation at all. We will call these
unphysical deformations\foot{These exist because of the presence of
non-trivial (0,1)-forms on the torus, and
can be seen more directly by noting that $h^{2,1} = 9$, but there are only 6
independent $\delta g_{ij}$}.

It can be shown that the flux superpotential \superpot\ in the $T^6/\IZ_2$
case \KST\ depends in
general on all the $\tau^{ij}$, even
those which are unphysical, evidence that the full supergravity description
of this
system \ferrara\ is more
complicated than the naive extrapolation of \noscale\ and \superpot\ to the
torus case.  To avoid this complication,
we turn to computing the quantities we will need from the dimensional
reduction of the ten dimensional
type IIB component Lagrangian.

\subsec{Dimensional Reduction}

We follow \GKP\ and \KST\ through most of this subsection. The Type IIB
supergravity action in ten dimensional
Einstein frame is \JoeTwo
\eqn\IIBsugra{\eqalign{S_{\rm IIB} &= {1\over2{\kappa_{10}}^2}\int
d^{10}x\sqrt{-g}\biggl(R-{\partial_M\phi~\partial^M\phi\over2({\rm Im} ~\phi)^2}-{G_{(3)}\cdot\bar G_{(3)}\over
2\cdot3!\cdot{\rm Im} ~\phi} -{\tilde F_{(5)}{}^2\over4\cdot 5!}\biggr)\cr&+{1\over2{\kappa_{10}}^2}\int
{C_{(4)}\wedge G_{(3)}\wedge\bar G_{(3)}\over 4 i ({\rm Im} ~\phi)} + S_{\rm local}.}}
where
\eqn\fielddefs{\phi=
C_{(0)}+i/g_s,\qquad G_{(3)}= F_{(3)}-\phi H_{(3)},}
\eqn\Ffive{ \tilde F_{(5)} = F_{(5)}-\half C_{(2)}\wedge H_{(3)}+\half
F_{(3)}\wedge B_{(2)},}
with $*\tilde F_{(5)} =\tilde F_{(5)}$, $F_{(3)}=dC_{(2)}$, and
$H_{(3)}=dB_{(2)}$.
If we compactify on a six dimensional compact manifold ${\cal M}_6$, the
Bianchi identity for the 5-form field strength is \eqn\bianchi{d\tilde
F_{(5)} = d*\tilde F_{(5)} =
H_{(3)}\wedge F_{(3)} + 2{\kappa_{10}}^2\mu_3 \rho_3^{\rm local}} where
$\rho_3^{\rm local}$ is the number
density of local D3-brane charge sources. This can be integrated over ${\cal
M}_6$
to give
\eqn\intbianchi{{1\over2{\kappa_{10}}^2\mu_3} \int_{{\cal M}_6}H_{(3)}\wedge
F_{(3)} + Q_3^{\rm local}=0.}

In the case of interest to us, the local sources of D3-brane charge are
D3-branes, with charge +1, and
orientifold 3-planes with charge -1/4. In the particular case of $T^6 /
\IZ_2$, there are $2^6$ O3-planes,
so the condition becomes \eqn\tadpole{{1\over2}{1\over (2 \pi)^4
(\alpha')^2} \int_{T^6}H_{(3)} \wedge F_{(3)} +
N_{\rm D3} + \ha N_{\rm O3'}=16.} Since we wish to avoid
$\overline{D3}$-branes, this gives the constraint
\eqn\tadconst{{1\over (2 \pi)^4 (\alpha')^2} \int_{T^6} H_{(3)} \wedge
F_{(3)} \leq 32.}

There are various types of exotic O3-planes that would contribute to
\tadpole, but we will impose constraints so
that these do not appear in our constructions. In particular, we will use
the quantization conditions
\eqn\quantization{{1\over(2\pi)^2\alpha'}\int_\gamma F_{(3)}=m_\gamma\in 2
\IZ,\qquad
{1\over(2\pi)^2\alpha'}\int_\gamma H_{(3)}=n_\gamma\in 2 \IZ,} so that
$F_{(3)}$ and $H_{(3)}$ still obey
the standard quantization conditions even though all three cycles are halved
in volume by the $\IZ_2$
action. This ensures that all of the O3-planes are of the standard type
\FreyHF.

A convenient basis for $H^3(T^6)$ is
\eqn\basis{\eqalign{
    \alpha_0 &= dx^1\wedge dx^2\wedge dx^3, \cr
    \alpha_{ij} &= {1\over2}\epsilon_{ilm}
                   dx^l\wedge dx^m\wedge dy^j,\quad 1\le i,j\le3, \cr
    \beta^{ij} &= -{1\over2}\epsilon_{jlm}
                   dy^l\wedge dy^m\wedge dx^i, \quad 1\le i,j\le3, \cr
    \beta_0 &= dy^1\wedge dy^2\wedge dy^3}
}
which satisfies
\eqn\symp{\int_{{\cal M}_6} \alpha_I \wedge \beta^J =
\delta^J_I.}
For compactification on
$T^6$, flux configurations of $F_{(3)}$ and $H_{(3)}$ take values in
$H^3(T^6,\IZ)$. As shown in \KST,
these configurations are all consistent with the orientifold action, with
the quantization-condition caveat
mentioned previously, so $F_{(3)}$ and $H_{(3)}$ can be generically expanded
as
\eqn\expthree{\eqalign{ {1 \over
(2\pi)^2 \alpha'}F_{(3)}
    &=a^0\alpha_0+a^{ij}\alpha_{ij}+b_{ij}\beta^{ij} + b_0\beta^0,\cr
{1 \over (2 \pi)^2 \alpha' } H_{(3)}
    &= c^0\alpha_0 + c^{ij}\alpha_{ij} + d_{ij}\beta^{ij} +d_0\beta^0
    }}
with $a^0,a^{ij},b_{ij},b_0,c^0,c^{ij},d_{ij},d_0 \in 2 \IZ$

For the reasons discussed in the previous subsection, we will study the
component potential energy
\eqn\realpot{V_{\rm real} = {1\over{24 \kappa_{10}^2 ({\rm Im} \rho)^3}}
\int_M d^6y g^{1/2}
{{G_{mnp} \bar G^{mnp}}\over{{\rm Im}~\phi}}-{i\over{4 \kappa_{10}^2 ({\rm
Im} ~\phi})({\rm Im} \rho)^3}
\int_M G_{(3)} \wedge \bar G_{(3)}.}
This comes from \IIBsugra\ after moving to four-dimensional Einstein frame,
taking into account the fact
that the second term is, by \tadpole, the energy of the local sources.

\subsec{The symmetric locus}

In order to simplify our task, we will now specialize to
an enhanced symmetry locus on the
moduli space of the $T^6$.  This allows us to first minimize the potential
within the
symmetric locus.  Since the enhanced symmetry guarantees no
tadpoles in the directions transverse to the symmetric locus, the potential
is already extremized in those
directions.  What remains is then to study the mass matrix in the transverse
directions to check for tachyons.

This simplification is crucial even for computer searches, which use a
steepest descent method. The minima we
for which we search are local minima with small ``drainage basins.''
Generically the number of attempts required
to find a minimum (even if we know it exists) will go as $(l'/l)^f$ where
$l$ is the scale of the basin, $l'>l$
the characteristic distance of the basin from the origin, and $f$ the number
of degrees of freedom. Furthermore,
each attempt takes longer as $f$ increases, so minimizing $f$ is quite
important.  By implementing a symmetry
constraint we reduce $f$ to the number of directions respecting the
symmetry. Of course, we have no guarantee that the symmetry-breaking
directions will be
tachyon-free.

In our case, will impose the symmetry generated by $R_1$:
$(x^1,x^2,x^3,y^1,y^2,y^3) \rightarrow
(-x^1,-x^2,x^3,-y^1,-y^2,y^3)$ and $R_2$: $(x^1,x^2,x^3,y^1,y^2,y^3)
\rightarrow (x^1,-x^2,-x^3,y^1,-y^2,-y^3)$.
Under this action, only diagonal $\tau^{ij}$, the $\tau^{ii}$, are
preserved, and the only invariant three-forms
are $\alpha_0$, $\alpha_{ii}$, $\beta^{ii}$, and $\beta^0$, limiting the
flux quantum numbers to
$a^0$, $a^{ii}$, $b_{ii}$, and $b_0$. If we limit ourselves to complex
structure moduli and the axion-dilaton,
this leaves us with eight real degrees of freedom, representing the three
diagonal tau and the axion-dilaton.

\subsec{Solutions}

Using the potential \realpot\ restricted to the symmetric locus, with fixed
$\rho$, we performed a randomized
search for nonzero local minima with the ``FindMinimum'' function of
Mathematica 4.1, which uses a modification
of Powell's method for minimization \mathematica. We tried more than $10^4$
different relatively prime choices
of flux quantum numbers $a,b,c,d \leq 10$, and for each flux choice, 50
attempts were made to find a nonzero
minimum. We found four distinct, but possibly related, nonzero minima--their
characteristics can be found in
Appendix A\foot{One could, at this point, consider these to be solutions in
the orientifold $T^6/ \IZ_2^3$
defined by the combined action. However, the flux quantization condition is
not clear in this case, and under
the most conservative choice, fluxes in $H^3(T^6, 8 \IZ)$, the condition
analogous to \tadconst\ is not
satisfied by any of our solutions.}.

We then performed the second step (checking for tachyons in the transverse
directions) by looking at the
derivatives of the full potential \realpot\ in terms of the full real metric
and the axion-dilaton. All first
derivatives were zero, which was expected for the symmetry-breaking K\"ahler
directions and all complex
structure directions. Interestingly, in all cases, almost all moduli were
fixed, with only three or four massless
directions remaining and no tachyons. These data are also in Appendix A.

\newsec{Conclusions}

We have exhibited tachyon-free models which satisfy \conditions.  In order to stabilize the three or four
remaining K\"ahler moduli, including the volume modulus, one would need to combine our contribution \volscale\
with non-perturbative and $\alpha^\prime$ contributions to the potential as in \KKLT. This provides a class of
models without the need for an antibrane in a warped throat and may facilitate more explicit model-building
analyses, since computations are easier in symmetric toroidal orientifolds of the sort we studied. We also
regard our success in finding explicit models in the toroidal case suggestive of their existence in the richer
context of generic Calabi-Yau manifolds. The uniformly non-negative mass-squares we encountered in the symmetry
breaking directions were a pleasant surprise, and it would be nice to find a simple analytical explanation for
that result.  The hierarchy of mass scales appearing in our results is also intriguing and may have interesting
applications.

\noindent{\bf Acknowledgements}

We thank Wu-yen Chuang for early collaboration and P. Candelas, D. Freed, S.
Kachru, X. Liu, L. McAllister, and M.
Schulz for many helpful discussions. We are supported in part by the DOE
under contract DE-AC03-76SF00515 and by
the NSF under contract 9870115.

\appendix{A}{Data}

All solutions have the same choices for 12 of the fluxes:
\eqn\commonflux{\eqalign{
      (b_0, b_{11}, b_{22}, b_{33}) &= (4,0,0,0) \cr
      (c^0, c^{11}, c^{22}, c^{33}) &= (4,-2,-2,-2) \cr
      (d_0, d_{11}, d_{22}, d_{33}) &= (-4,-2,-2,-2). \cr
}}

The locations and first derivatives of all vacua were calculated to at least
12 significant figures, though only
a few are given here, and all calculated values claimed to be zero were
smaller than $10^{-12}$ and of the same
order as other calculational artifacts. In the first table, the integer
under D3 is the number of D3-branes
required to satisfy \tadpole.

\bigskip
{\offinterlineskip \tabskip=0pt \halign{ \vrule height2.75ex depth1.25ex
width 0.6pt #\tabskip=1em & \hfil {\rm
#} \hfil &\vrule # & \hfil # \hfil  & \hfil # \hfil & \hfil # \hfil & \hfil
# \hfil &\vrule # & \hfil # \hfil
&\vrule # & \hfil # \hfil &#\vrule width 0.6pt \tabskip=0pt\cr
\noalign{\hrule height 0.6pt} &\omit Vacuum
&&\omit $a^0$ & \omit $a^{11}$ & \omit $a^{22}$ & \omit $a^{33}$ & &\omit
D3s && \omit $V \cdot (\alpha^\prime)^2
(\rho-\bar\rho)^3$ &\cr
\noalign{\hrule} & A && 4 & 2 & 0 & -2 && 0 && $7.424 \times 10^{-6}$ &\cr
\noalign{\hrule} & B && 4 & 0 & -2 & -2 && 4 && $1.029 \times 10^{-4}$ &\cr
\noalign{\hrule} & C && 4 & 0 & 0 & -2 && 2 && $3.609 \times 10^{-5}$ &\cr
\noalign{\hrule} & D && 4 & -2 & -2 & 2 && 2 && $4.233 \times 10^{-5}$ &\cr
\noalign{\hrule height 0.6pt} }}
\medskip

\medskip
{\offinterlineskip \tabskip=0pt \halign{ \vrule height2.75ex depth1.25ex
width 0.6pt #\tabskip=1em & \hfil {\rm
#} \hfil &\vrule # & \hfil # \hfil  & \hfil # \hfil & \hfil # \hfil & \hfil
# \hfil &#\vrule width 0.6pt
\tabskip=0pt\cr \noalign{\hrule height 0.6pt} &\omit Vac. && $\tau^{11}$ &
$\tau^{22}$ & $\tau^{33}$ & $\phi$
&\cr \noalign{\hrule} & A && 2.653 - 1.851 i & 2.032 - 0.4752 i & 0.2534 -
0.4578 i & 0.9022 + 0.8783 i &\cr
\noalign{\hrule} & B && 1.477 - 1.427 i & 0.6650 - 0.5250 i & 0.6650 -
0.5250 i & 1.177 + 0.4567 i &\cr
\noalign{\hrule} & C && 1.523 - 0.7046 i & 1.523 - 0.7046 i & 0.4489 -
0.4800 i & 1.161 + 0.5285 i &\cr
\noalign{\hrule} & D && 0.5846 - 0.4487 i & 0.5846 - 0.4487 i & 1.225 -
2.835 i & 0.8017 + 1.108 i &\cr
\noalign{\hrule height 0.6pt} }}
\bigskip
\medskip

The following tables show the eigenvalues of the mass matrix at the minima,
in the 23 real directions of all
metric moduli and the axion-dilaton.

\bigskip
{\offinterlineskip \tabskip=0pt \halign{ \vrule height2.75ex depth1.25ex
width 0.6pt #\tabskip=1em & \hfil {\rm
#} \hfil &\vrule # & \hfil # \hfil  & \hfil # \hfil & \hfil # \hfil & \hfil
# \hfil & \hfil # \hfil &#\vrule
width 0.6pt \tabskip=0pt\cr \noalign{\hrule height 0.6pt} &\omit Vac. &&&
\omit Mass matrix eigenvalues of
$V \cdot (\alpha^\prime)^2
(\rho-\bar\rho)^3$ \span \span &&\cr
\noalign{\hrule} & A && 0.4404 & 0.09416 & 0.06495 & 0.03706 & 0.004580 &\cr
      &&& 0.002429 & 0.001306 & 0.001286 & $7.215 \times 10^{-4}$ & $6.756
\times 10^{-4}$ &\cr
      &&& $6.192 \times 10^{-4}$ & $3.862 \times 10^{-4}$ & $3.652 \times
10^{-4}$ & $3.635 \times 10^{-4}$ & $2.996 \times 10^{-4}$ &\cr
      &&& $1.831 \times 10^{-4}$ & $1.597 \times 10^{-4}$ & $1.487 \times
10^{-4}$ & $9.190 \times 10^{-6}$ & $1.802 \times 10^{-7}$ &\cr
      &&& 0 & 0 & 0 &&&\cr
\noalign{\hrule} & B && 0.02050 & 0.01523 & 0.01523 & 0.008724 & 0.008334
&\cr
      &&& 0.007821 & 0.007525 & 0.0023505 & 0.001444 & $9.480 \times
10^{-4}$
&\cr
      &&& $9.480 \times 10^{-4}$ & $2.846 \times 10^{-4}$ & $1.781 \times
10^{-4}$ &  $1.430 \times 10^{-4}$ & $1.074 \times 10^{-4}$ &\cr
      &&& $1.074 \times 10^{-4}$ & $7.303 \times 10^{-5}$ & $2.791 \times
10^{-5}$ & $2.791 \times 10^{-5}$ & 0 &\cr
      &&& 0 & 0 & 0 &&&\cr
\noalign{\hrule height 0.6pt} }}
\medskip

{\offinterlineskip \tabskip=0pt \halign{ \vrule height2.75ex depth1.25ex
width 0.6pt #\tabskip=1em & \hfil {\rm
#} \hfil &\vrule # & \hfil # \hfil  & \hfil # \hfil & \hfil # \hfil & \hfil
# \hfil & \hfil # \hfil &#\vrule
width 0.6pt \tabskip=0pt\cr \noalign{\hrule height 0.6pt} &\omit Vac. &&&
\omit Mass matrix eigenvalues of
$V \cdot (\alpha^\prime)^2
(\rho-\bar\rho)^3$ \span \span &&\cr
\noalign{\hrule} & C && 0.05010 & 0.02131 & 0.02131 & 0.01151 & 0.002869
&\cr
      &&& 0.001775 & 0.001629 & 0.001401 & 0.001401 & 0.001258 &\cr
      &&& $7.722 \times 10^{-4}$ & $6.701 \times 10^{-4}$ & $3.245 \times
10^{-4}$ & $2.281 \times 10^{-4}$ & $2.281 \times 10^{-4}$ &\cr
      &&& $1.679 \times 10^{-4}$ & $1.080 \times 10^{-4}$ & $1.080 \times
10^{-4}$ & $3.813 \times 10^{-5}$ & 0 &\cr
      &&& 0 & 0 & 0 &&&\cr
\noalign{\hrule} & D && 0.07233 & 0.07233 & 0.07027 & 0.01616 & 0.01279 &\cr
      &&& 0.008878 & 0.007767 & 0.005338 & 0.002045 & 0.001176 &\cr
      &&& $8.341 \times 10^{-4}$ & $8.341 \times 10^{-4}$ & $3.082 \times
10^{-4}$ & $3.082 \times 10^{-4}$ & $1.971 \times 10^{-4}$ &\cr
      &&& $1.098 \times 10^{-4}$ & $9.928 \times 10^{-5}$ & $3.293 \times
10^{-5}$ & $3.293 \times 10^{-5}$ & 0 &\cr
      &&& 0 & 0 & 0 &&&\cr
\noalign{\hrule height 0.6pt} }}
\medskip

\listrefs

\end